\documentstyle[aps,12pt,preprint,psfig]{revtex}
\tightenlines

\def \to {\rightarrow}
\def \beq {\begin{equation}}
\def \eeq {\end{equation}}
\def \ba {\begin{eqnarray}}
\def \ea {\end{eqnarray}}

\def \< {\left <}
\def \> {\right >}
\begin{document}
\preprint{
\vbox{
\halign{&##\hfil\cr
        & PKU-TP-98-31\cr
        & hep-ph/yymmnn \cr
        & June 1998 \cr\cr}}
        }
\baselineskip 18pt
\renewcommand{\thesection}{\Roman{section}}
\draft
 
\title{$D$-wave heavy quarkonium production in fixed target experiments}

\author{Feng Yuan and Cong-Feng Qiao}
\address{\small {\it Department of Physics, Peking University, Beijing
100871, People's Republic of China}}
\author{Kuang-Ta Chao}
\address{\small {\it China Center of Advanced Science and Technology
(World Laboratory), Beijing 100080, People's Republic of China\\
and Department of Physics, Peking University, Beijing 100871,
People's Republic of China}}

\maketitle
\begin{abstract}

We calculate the $D$-wave heavy quarkonium production at fixed target 
experiments under the NRQCD factorization formalism. We find that the color 
octet contributions are two orders of magnitude larger than color-singlet
contributions if color-octet matrix elements are taken according to  
the NRQCD velocity scaling rules. Within the theoretical uncertainties,
the prediction for the production rate
of $2^{--}$ $D$-wave charmonium state agrees with the preliminary result
of E705 and other experiments.
Searching for the $1^{--}$ $D$-wave state $\psi(3770)$ is further suggested.
\end{abstract}

\pacs{PACS number(s): 12.40.Nn, 13.85.Ni, 14.40.Gx}

Studies of heavy quarkonium production in high energy collisions provide
important information on both perturbative and nonperturbative QCD.
Recent progress in the this area was stimulated by the experiment results of 
{\bf CDF} at the Fermilab Tevatron. In the 1992-1993 run, the 
{\bf CDF} data \cite{cdf} for the prompt production rates of $\psi$ and
$\psi^\prime$ at large transverse momentum were observed to be orders of
magnitude larger than the lowest order perturbative calculation based on 
the color-singlet model. At the same time, a new framework in treating
quarkonium production and decays has been advocated by Bodwin, Braaten
and Lepage in the context of nonrelativistic quantum chromodynamics
(NRQCD) \cite{bbl}. In this approach, the production process is
factorized into short and long distance parts, while the latter is
associated with the nonperturbative matrix elements of four-fermion
operators. This factorization formalism gives rise to a new production
mechanism called the color-octet mechanism, in which the heavy-quark and
antiquark pair is produced at short distance in a color-octet configuration
and subsequently evolves nonperturbatively into physical quarkonium state.
The color-octet terms in the gluon fragmentation to $J/\psi$($\psi'$) have been 
considered to explain the $J/\psi$($\psi'$) surplus problems discovered
by CDF \cite{surplus,s1}. Taking the nonperturbative $\< {\cal
O}^{J/\psi}_8(^3S_1)\> $ and $\< {\cal O}^{\psi'}_8(^3S_1)\> $ as input
parameters, the CDF surplus problem for $J/\psi$ and $\psi'$ can be explained
as the contributions of color-octet terms due to gluon fragmentation.
In the past few years, applications of the NRQCD factorization formalism
to $J/\psi$($\psi^\prime$) production at various experimental facilities
have been studied \cite{annrev}.

Even though the color-octet mechanism has gained some successes in describing
the production and decays of heavy quark bound systems, it still needs more
effort to go before finally setting its position and role in heavy quarkonium
physics. (For instance, the photoproduction data from HERA \cite{photo} puts
a question about the universality of the color-octet matrix elements
\cite{photo2}, in which the fitted values of the matrix elements 
$\langle {\cal O}^{J/\psi}_8 ({}^1S_0)\rangle$ and $\langle {\cal
O}^{J/\psi}_8({}^3P_J)\rangle$ are one order of magnitude smaller than
those determined from the Tevatron data \cite{s1})\footnote{Possible
solutions for this problem have been suggested recently in\cite{explain}.}
So, we must find
other processes to test the color-octet mechanism in the heavy quarkonium
production. 

In our previous studies \cite{qiao,gf,z0,bdw}, we propose the $D$-wave
heavy quarkonia production as a crucial test for the color-octet mechanism. 
We have calculated the $D$-wave quarkonium production via gluon fragmentation
at the hadron colliders\cite{gf}, in the $Z^0$ decays \cite{z0}, and the
$D$-wave charmonia production in $B$ decays \cite{bdw}. All these results
show that the color-octet mechanism is crucially important to $D$-wave
charmonium production, and the color-octet contributions are found to be over
two orders of magnitude larger than the color-singlet contributions. 

In this paper, we analyze the $D$-wave heavy quarkonium production in
fixed target experiments, at energies in which the color-octet gluon
fragmentation is not dominant. However, the analysis of $S$-wave
charmonium and bottomonium production in fixed target experiments show
that the color-octet contribution is important in removing large
discrepancies between experiments and predictions from the
color-singlet model\cite{fixed-s}.  We can further expect that the
color-octet contribution may be dominant in the production of
$D$-wave quarkonium production in fixed target, because within the
color-singlet model the $D$-wave heavy quarkonia production is always
suppressed by the second derivative of the wave function at the origin.
But this suppression can be avoided in the color-octet model.
Therefore, the $D$-wave heavy quarkonium production
may provide another test for the color-octet mechanism.
In addition, on the experimental side,
there are some clues for the $D$-wave $2^{--}$ charmonium state in $E705$
$300$ GeV $\pi^\pm$- and proton-Li interaction experiment\cite{14p}. In
this experiment there is an abnormal phenomenon that in the
$J/\psi\pi^+\pi^-$ mass spectrum, two peaks at $\psi(3686)$ and at $3.836$
GeV (given to be the $2^{--}$ state) are observed and they have almost the
same height. Obviously, this situation is difficult to explain based upon
the color-singlet model. However, it might be explained within the scope
of NRQCD.

In NRQCD the Fock state expansion for
${}^3D_J$ states is 
\begin{eqnarray} 
\label{zankai} 
|{}^3D_J \rangle
=O(1)|Q\bar{Q}({}^3D_{J},\b{1}) \rangle +
O(v)|Q\bar{Q}({}^3P_{J'},\b{8})g\rangle +
    O(v^2)|Q\bar{Q}({}^3S_1,\b{8}~ or~ \b{1})gg\rangle +\cdots. 
\end{eqnarray} 
In the above expansion, the contributions to the NRQCD matrix elements for
the production of the $D$-wave charmonium from the three terms are the same
order of $v$ according to the NRQCD velocity scaling rules. So, in the
factorization formula of $D$-wave quarkonium production, all of these
three terms should be considered. Accordingly, the production cross section
for the physical $D$-wave quarkonium state $\delta_J$ in hadron process \footnote{Here, 
the symbol $\delta_J$ denotes the physical spin-triplet $D$-wave heavy
quarkonium state. The notation ${}^{2S+1}D_J$ represents the $c\bar c$ or
$b\bar b$ pair configurations with angular momentum $L=2$.}
\beq 
A+B\to \delta_J + X 
\eeq 
can be written as 
\beq \label{xs}
\sigma(A+B\to \delta_J +X)=\sum \limits_{i,j}\int\limits_0^1
dx_1dx_2f_{i/A}
        (x_1)f_{j/B}(x_2)\hat{\sigma}(ij\to\delta_J), 
\eeq        
\beq
\hat\sigma(ij\to \delta_J)=\sum\limits_n F(ij\to n)\< {\cal
O}_n^{\delta_J}\> .
\eeq
Here, $n$ denotes $c\bar c$ pair configuration in the three Fock states of 
Eq. (\ref{zankai}) (including angular momentum $^{2S+1}L_J$ and color index
1 or 8). $F(ij\to n)$ is the short distance coefficient for the subprocess
$ij\to n$. $\< {\cal O}_n^{\delta_J}\> $ is the long distance
non-perturbative matrix element which represents the property of the
$\bar c c$ pair in $n$ configuration evolving into the physical state
$\delta_J$. The short distance coefficient $F$ can be calculated by using
perturbative QCD in expansion of powers of $\alpha_s$. The long distance
matrix elements are still not available from the first principle at
present, which can are obtained by fitting the theoretical prediction with
the experimental result in literatures. By the naive NRQCD velocity
scaling rules, these three matrix elements are of 
the same order in powers of $v$. Their scaling properties are,
\ba
\< {\cal O}_1^{\delta_J}({}^3D_J)\> \sim M_c^7 v^7,~~
\< {\cal O}_8^{\delta_J}({}^3S_1)\> \sim M_c^3 v^7,~~
\< {\cal O}_8^{\delta_J}({}^3P_J')\> \sim M_c^5 v^7.
\ea
That is to say that the color-octet and color-singlet processes are of the
same order in powers of $v$. However, the color-octet processes are
enhanced by an order of $\alpha_s$ over the color-singlet processes.
So, in the hadroproduction of $D$-wave heavy quarkonium at fixed target 
experiments the color-octet contributions should be included.

First, we will calculate the color-singlet contribution to
${}^3D_J$ quarkonium hadroproduction. The leading order color-singlet
contribution comes from the gluon-gluon fusion process $gg\to {}^3D_J g$.
The production rate of this process can be calculated by making use of the
covariant formalism. The general form of the wave function of the
spin-triplet heavy quarkonium state (with angular momentum ${}^3L_J$)
can be written as
\begin{equation}
 \Phi(P,\vec{q})=\frac{1}{M}\sum_{s m}
 \langle JM|1s Lm\rangle \Lambda^1_+ (\vec{p_1})\gamma_0 
  \not\!{\epsilon^{(s)}}(M+\not\!{P})\gamma_0\Lambda^2_- (\vec{p_2})\psi_{Lm}
         (\vec{q}),
\end{equation}
where $\epsilon^{(s)}$ is the polarization vector associated with the spin-triplet 
states. $\Lambda^1_+(\vec{p_1})$ and $\Lambda^2_-(\vec{p_2})$ 
are positive energy projection operators of quark and antiquark .
\begin{eqnarray}
  \label{add8}
  \Lambda^1_+(\vec{p_1})=\frac{E_1+\gamma_0 \vec{\gamma}\cdot\vec{p_1} + 
   M_c\gamma_0}{2E_1},~~~~
  \Lambda^2_-(\vec{p_2})=\frac{E_2-\gamma_0 \vec{\gamma}\cdot\vec{p_2} - 
  M_c\gamma_0}{2E_2}.
\end{eqnarray}

For the $D$-wave function, $\Phi(P,\vec{q})$ must be expanded to the second
order in the relative momentum $\vec{q}$. The first and second derivatives
of the wave function are
\begin{equation}
\Phi_\alpha(\vec{q})=\frac{-1}{2 M_c^2 M}\sum_{sm}\langle JM|1sLm\rangle  
[M_c\gamma_\alpha \!\not{\epsilon^{(s)}}(M+\not\!{P}) + 
M_c\!\not{\epsilon^{(s)}}(M+\not\!{P})\gamma_\alpha ]
\psi_{Lm}(\vec{q});
\end{equation}
\begin{equation}
\Phi_{\alpha\beta}(\vec{q})=\frac{-1}{2 M_c^2 M}\sum_{sm}\langle
JM|1sLm\rangle \gamma_\alpha\!\not{\epsilon^{(s)}}(M+\not\!{P}) 
\gamma_\beta\psi_{Lm}(\vec{q}).
\end{equation}
After integrating over the relative momentum $\vec{q}$, the orbit angular
momentum part of the wave function will depend on the radial wave function
or its derivatives at the origin $R_S(0)$, $R_P^\prime(0)$, and
$R_D^{\prime\prime}(0)$, respectively, for $S$-wave ($L=0$), $P$-wave
($L=1$), and $D$-wave ($L=2$) states,
\begin{equation}
 \int\frac{d^3 q}{(2\pi)^3}\psi_{00}({\bf q})=\frac{1}{\sqrt{4\pi}}R_S(0),
\end{equation}
\begin{equation}
 \int\frac{d^3 q}{(2\pi)^3}q^{\alpha}
 \psi_{1m}({\bf q})=i\epsilon^{\alpha}\sqrt{\frac{3}{4\pi}}R_P^{\prime}(0),
\end{equation}
\begin{equation}
 \int\frac{d^3 q}{(2\pi)^3}q^{\alpha}q^{\beta}
 \psi_{2m}({\bf q})=e^{\alpha\beta}_m\sqrt{\frac{15}{8\pi}}
 R_D^{\prime\prime}(0),
\end{equation}
where the polarization tensor's label $m$ is magnetic quantum number. For 
the spin-triplet case where $J=1,~2,~3$, using explicit Clebsch-Gordan
coefficients, there are following relations for the three cases \cite{be}.
\begin{eqnarray}
\nonumber  
\sum_{sm}\langle 1J_z|1s2m\rangle\epsilon_{\alpha\beta}^{(m)} 
\epsilon_\rho^{(s)}
&=&-[{3\over 20}]^{1/2}[(g_{\alpha\rho}-\frac{p_\alpha
p_\rho}{4M_c^2})\epsilon_\beta^{(J_z)}+(g_{\beta\rho}-\frac{p_\beta
p_\rho}{4M_c^2})\epsilon_\alpha^{(J_z)}\\
&~&-{2\over 3}(g_{\alpha\beta}-\frac{p_\alpha
p_\beta}{4M_c^2})\epsilon_\rho^{(J_z)}],\\
\sum_{sm}\langle 2J_z|1s2m\rangle\epsilon_{\alpha\beta}^{(m)}
\epsilon_\rho^{(s)}
&=&\frac{i}{2\sqrt{6}M_c}(\epsilon_{\alpha\sigma}^{(J_z)}
\epsilon_{\tau\beta\rho\sigma^{\prime}}p^\tau g^{\sigma\sigma^\prime}+
\epsilon_{\beta\sigma}^{(J_z)}\epsilon_{\tau\alpha\rho\sigma^{\prime}}
p^\tau g^{\sigma\sigma^\prime}),\\
\sum_{sm}\langle 3J_z|1s2m\rangle\epsilon_{\alpha\beta}^{(m)}
\epsilon_\rho^{(s)}
&=&\epsilon_{\alpha\beta\rho}^{(J_z)}.
\end{eqnarray}
Here, $\epsilon_\alpha$, $\epsilon_{\alpha\beta}$,
$\epsilon_{\alpha\beta\rho}$
are the spin-one, spin-two and spin-three polarization tensors which obey the
projection relations \cite{be}
\begin{eqnarray}
\sum_m \epsilon_\alpha^{(m)}\epsilon_\beta^{(m)}
&=&(-g_{\alpha\beta}+\frac{p_\alpha p_\beta}{4M_c^2})
\equiv{\cal P}_{\alpha\beta},\\
\sum_m \epsilon_{\alpha\beta}^{(m)}\epsilon_{\alpha^\prime\beta^\prime}^{(m)}
&=&{1\over 2}[{\cal P}_{\alpha\alpha^\prime}{\cal P}_{\beta\beta^\prime}+
{\cal P}_{\alpha\beta^\prime}{\cal P}_{\beta\alpha^\prime}]
-{1\over 3}{\cal P}_{\alpha\beta}{\cal P}_{\alpha^\prime\beta^\prime},\\
\nonumber
\sum_m \epsilon_{\alpha\beta\rho}^{(m)}
\epsilon_{\alpha^\prime\beta^\prime\rho^\prime}^{(m)}  
&=&{1\over 6}({\cal P}_{\alpha\alpha^\prime}
{\cal P}_{\beta\beta^\prime}{\cal P}_{\rho\rho^\prime}
+{\cal P}_{\alpha\alpha^\prime}{\cal P}_{\beta\rho^\prime}
{\cal P}_{\beta\rho^\prime}
+{\cal P}_{\alpha\beta^\prime}{\cal P}_{\beta\alpha^\prime}
{\cal P}_{\rho\rho^\prime}\\
\nonumber
&~&~~~~+{\cal P}_{\alpha\beta^\prime}{\cal P}_{\beta\rho^\prime}
{\cal P}_{\rho\alpha^\prime}                            
+{\cal P}_{\alpha\rho^\prime}{\cal P}_{\beta\beta^\prime}
{\cal P}_{\rho\alpha^\prime}
+{\cal P}_{\alpha\rho^\prime}{\cal P}_{\beta\alpha^\prime}
{\cal P}_{\rho\beta^\prime})\\
\nonumber
&-&{1\over 15}({\cal P}_{\alpha\beta}
{\cal P}_{\rho\alpha^\prime}{\cal P}_{\beta^\prime\rho^\prime}
+{\cal P}_{\alpha\beta}{\cal P}_{\rho\beta^\prime}
{\cal P}_{\alpha^\prime\rho^\prime}
+{\cal P}_{\alpha\beta}{\cal P}_{\rho\rho^\prime}
{\cal P}_{\alpha^\prime\beta^\prime}\\
\nonumber
&~&~~~~+{\cal P}_{\alpha\rho}{\cal P}_{\beta\alpha^\prime}
{\cal P}_{\beta^\prime\rho^\prime}
+{\cal P}_{\alpha\rho}{\cal P}_{\beta\beta^\prime}
{\cal P}_{\alpha^\prime\rho^\prime}
+{\cal P}_{\alpha\rho}{\cal P}_{\beta\rho^\prime}
{\cal P}_{\alpha^\prime\beta^\prime}\\               
&~&~~~~+{\cal P}_{\beta\rho}{\cal P}_{\alpha\alpha^\prime}
{\cal P}_{\beta^\prime\rho^\prime}
+{\cal P}_{\beta\rho}{\cal P}_{\alpha\beta^\prime}
{\cal P}_{\alpha^\prime\rho^\prime}
+{\cal P}_{\beta\rho}{\cal P}_{\alpha\rho^\prime}
{\cal P}_{\alpha^\prime\beta^\prime}).
\end{eqnarray}
So, the differential cross section for the color-singlet process 
$gg\to \delta_JX$ can be calculated, as an expansion of the form in
Eq.(\ref{xs}),
\beq
\label{xs-cs}
\frac{d\sigma(gg\to \delta_Jg)_1}{dt}=F(gg\to {}^3D_Jg)\times \< {\cal O}
_1^{\delta_J}({}^3D_J)\> ,                          
\eeq
where the color-singlet matrix element $\< {\cal O}
_1^{\delta_J}({}^3D_J)\> $
can be related to the second derivative of the nonrelativistic radial wave
function at the origin $|R_D^{\prime\prime}(0)|^2$ for $D$-wave by
\ba
\< {\cal O}_1^{\delta_J}({}^3D_J)\>
&=&\frac{15(2J+1)N_c}{4\pi}|R_D^{\prime\prime}(0)|^2.
\ea
The calculation of the short distance coefficient $F(gg\to {}^3D_J)g$
is straightforward. Because the results are lengthy, we give the
expressions in the Appendix. As in the case of gluon fragmentation
to color-singlet ${}^3D_J$\cite{gf}, $gg\to {}^3D_Jg$ processes also have
the infrared divergences involved, which are associated with the soft
gluon in the final state. In our calculations of the cross section for
these processes, we follow the way in Ref.\cite{gf} by imposing a lower
cutoff $\Lambda$ on the energy of the outgoing gluon in the quarkonium
rest frame. The cutoff $\Lambda$ can be set to be $m_Q$ to avoid large
logarithms in the divergence terms.

For the color-octet contributions in the $gg$, $gq$, $q\bar q$
subprocesses, we readily have\cite{s1,fixed-s}
\ba
\hat{\sigma}(gg\to \delta_J)_8&=&\frac{5\pi^3\alpha_s^2}{12(2m_Q)^3s}
\delta(x_1x_2-4m_Q^2/s)[\frac{3}{m_Q^2}\< {\cal O}_8^{\delta_J}({}^3P_0)\> 
+\frac{4}{5m_Q^2}\< {\cal O} _8^{\delta_J} ({}^3P_2)\> ],\\
\hat{\sigma}(gq\to \delta_J)_8&=&0,\\
\hat{\sigma}(q\bar q\to \delta_J)_8&=&\frac{16\pi^3\alpha_s^2}{27(2m_Q)^3s}
        \delta(x_1x_2-4m_Q^2/s)\< {\cal O}_8^{\delta_J}({}^3S_1)\> .
\ea
Here $\sqrt{s}$ is the center-of-mass energy, and $\alpha_s$ is normalized at 
the scale $2m_Q$.

Our numerical results are displayed in Fig.1 to Fig.4. We use the 
Gl\"uck-Reya-Vogt (GRV) LO parton distributions for the proton and
the pion \cite{grv1,grv2}. We set the renormalization scale to be $2m_Q$.

In Fig.1 and Fig.2, we plot the total cross section of $D$-wave heavy
quarkonium for proton-nucleon collisions for $x_F>0$, where we choose the
color-octet matrix elements to be related to the color-singlet matrix elements 
according to the naive NRQCD velocity scaling rules,
\ba
\frac{\< {\cal O}_8^{\delta^c_1}({}^3S_1)\> }{M_c^3}&\approx &
        \frac{\< {\cal O}_8^{\delta^c_1}({}^3P_1)\> }{M_c^5}\approx
        \frac{\< {\cal O}_1^{\delta^c_1}({}^3D_1)\> }
        {M_c^7},\\
\frac{\< {\cal O}_8^{\delta^b_1}({}^3S_1)\> }{M_b^3}&\approx &
        \frac{\< {\cal O}_8^{\delta^b_1}({}^3P_1)\> }{M_b^5}\approx
        \frac{\< {\cal O}_1^{\delta^b_1}({}^3D_1)\> }
        {M_b^7}.
\ea
And the heavy quark spin symmetry relations for the color-octet matrix 
elements have been used,
\ba
\< {\cal O}_8^{\delta_J}({}^3S_1)\> &\approx &\frac{2J+1}{3}
        \< {\cal O}_8^{\delta_1}({}^3S_1)\> ,\\
\< {\cal O}_8^{\delta_J}({}^3P_1)\> &\approx &\frac{2J+1}{3}
        \< {\cal O}_8^{\delta_1}({}^3P_1)\> .
\ea
The values of color-singlet matrix elements are obtained from the potential
model calculation, 
$|R_D^{\prime\prime}(0)|_c^2=0.015 GeV^7$,
$|R_D^{\prime\prime}(0)|_b^2=0.637 GeV^7$\cite{potential}.
Fig.1 is the cross section for the $D$-wave charmonium and Fig.2 for the $D$-wave bottomonium
respectively. From these figures, we can see that the color-octet
contributions are two orders of magnitude larger than the color-singlet
contributions.

It must be noted that all of the cross sections are expressed in terms
of heavy quark mass as $m_c$ and $m_b$, for which we take values as
$$m_c=1.5GeV,~~~M_b=4.9 GeV.$$ If we choose the charm quark mass as one half
of the charmonium mass (e.g.,
$\delta^c\approx 3.8GeV$), the theoretical prediction
of the cross section will be reduced by an order of magnitude for the
color-singlet contributions and by a factor of five for the color-octet
contributions at a typical energy scale $\sqrt{s}=24GeV$.
Compared with this large uncertainty, the variation due to the choice of 
parton distribution functions and $\alpha_s(\mu)$ is negligible.

In Fig.3 and Fig.4, we plot the cross sections of $D$-wave charmonium
and bottomonium states in pion-nucleon collisions for $x_F>0$.
Similar to the case of proton-nucleon collisions, in pion-nucleon
collisions we can see from these figures that the color-octet
contributions are two order of magnitude larger than the color-singlet
contributions.

Among the three triplet states of $D$-wave charmonium, $\delta^c_2$ is the
most promising candidate to discover firstly. Its mass falls in the range
of $3.810\sim 3.840$ GeV in the potential model calculation, that is
below the $D{\bar D}^*$ threshold but above the $D{\bar D}$ threshold. 
However, the parity conservation forbids it decaying into $D\bar D$. 
It, therefore, is a narrow resonance. Its main decay modes are expected to be,
\begin{equation}
\delta^c_2 \rightarrow J/\psi\pi\pi,~~~\delta^c_2\rightarrow {}^3P_J
\gamma(J=1,2),~~~
\delta^c_2\rightarrow 3g.
\end{equation}
The hadronic transition rate of $\delta^c_2\rightarrow J/\psi\pi^+\pi^-$
is estimated to be \cite{z0}
\begin{equation}
\label{y1}
\Gamma(\delta^c_2 \rightarrow J/\psi \pi^+\pi^-)=\Gamma(^3D_1\rightarrow
J/\psi \pi^+\pi^-) \approx 46~ keV.
\end{equation}
For the E1 transition $\delta_c^2\rightarrow {}^3P_J\gamma(J=1,2)$, using
the potential model with relativistic effects being  considered
\cite{dchao}, we find
\begin{equation}
\label{y2}
\Gamma(\delta^c_2 \rightarrow \chi_{c1}\gamma)=250~ keV,~~~
\Gamma(\delta^c_2 \rightarrow \chi_{c2}\gamma)=60~ keV,
\end{equation}
where the mass of $\delta^c_2$ is set to be $3.84 GeV$. 
As for the $\delta^c_2\rightarrow
3g$ annihilation decay, an estimate gives\cite{be}
\begin{equation}
\label{y3}
\Gamma(\delta^c_2\rightarrow 3g)=12~ keV.
\end{equation}
From (\ref{y1}), (\ref{y2}), and (\ref{y3}), we find
\begin{eqnarray}
\nonumber
\Gamma_{tot}(\delta^c_2)&\approx & \Gamma(\delta^c_2\rightarrow J/\psi
\pi\pi)
+\Gamma(\delta^c_2 \rightarrow \chi_{c1}\gamma)+\Gamma(\delta^c_2
\rightarrow \chi_{c2}\gamma) +\Gamma(\delta^c_2\rightarrow 3g)\\
&\approx &390~ keV,
\end{eqnarray}
and
\begin{equation}
\label{e17}
B(\delta^c_2\rightarrow J/\psi \pi^+ \pi^{-} )\approx 0.12.
\end{equation}
Comparing (\ref{e17}) with $B(\psi^\prime\rightarrow J/\psi
\pi^+\pi^-)=0.324\pm 0.026$,
the branching ratio of $\delta^c_2\rightarrow J/\psi \pi^+\pi^-$ is only
smaller by a factor of 3, and therefore the decay mode of
$\delta^c_2\rightarrow J/\psi \pi^+\pi^-$ could be observable in various
experiments, such as at hadron colliders and at fixed target experiments.

Multiplied by the assumed branching ratio of $\delta^c_2\rightarrow
J/\psi \pi^+ \pi^{-} $ above, we can estimate the production rate of
$\delta^c_2$ in pion-nucleon collisions. From Fig.3, at $\sqrt{s}\approx 24
GeV$, the cross section is predicted to be $\sigma(\pi^-N\to 
\delta^c_2+X)B(\delta^c_2\rightarrow J/\psi \pi^+ \pi^{-})=
4{\rm nb}$ per nucleon. This value is in agreement with the preliminary
result of E705, where the cross section is estimated to be
$5.3\pm 1.9\pm 1.3 nb$\cite{14p}. From Fig.3, we can see that the experimental
result of E705 collaboration can not be explained within the color-singlet
model, because the prediction of the color-singlet model is two orders of
magnitude smaller than the color-octet model.
This indicates that though the agreement with the E705 data can not be
taken too seriously (due to large theoretical uncertainties, e.g., the
choice of charm quark mass; the naive estimate for the color-octet matrix
elements), an appreciable $D$-wave signal can only be explained by the
color-octet mechanism.
Experimentally, the strong signal
of $J/\psi\pi^+\pi^-$ at $3.836 GeV$ observed by $E705$ is now questioned
by other experiments.
(E705 observed ($77\pm 21$) and ($58\pm 21$) background-subtracted events at $\psi'(2S)$
and 3.836 GeV respectively [14]; while E672-E706 reported ($224\pm 44$) and
($52\pm 30$) background-subtracted events at $\psi'(2S)$ and 3.836 GeV
respectively [20]).
Nevertheless, if the $E705$ result is
confirmed (even with a smaller rate, say, by a factor of 3, for the
signal at $3.836 GeV$, as might be implied by other experiments\cite{e627}),
the color-octet production mechanism may
provide a quite unique explanation for the $D$-wave charmonium
production. 

Moreover, the $1^{--}$ $D$-wave charmonium state $\psi(3770)$ may also be
observable at the fixed target experiments, with a rate smaller by a factor
of $5\over 3$ than the $2^{--}$ state, provided the decay channel $\psi(3770)
\to D\bar D$ (with about $100\%$ decay branching ratio to the charmed meson
pairs) can be detected.

In conclusion, in this paper, we have calculated the $D$-wave heavy
quarkonium production in fixed target experiments. We find that the
color-octet mechanism plays an important role in the production. The
color-octet contributions are two order of magnitude higher than the
color-singlet contributions both in proton-nucleon and pion-nucleon
collisions. Despite of some theoretical uncertainties, 
the prediction of the color-octet model is found to be in 
agreement with the preliminary result of E705 collaboration.
This may provide another positive support to the color-octet
production mechanism of heavy quarkonium.

\vskip 1cm

\begin{center}
{\bf {\large {Acknowledgments}\ }}
\end{center}

This work was supported in part by the National Natural Science Foundation
of China, the State Education Commission of China, and the State
Commission of Science and Technology of China.

\newpage
\appendix
\section*{}
The short distance coefficients for the color-singlet processes $gg\to
{}^3D_J$ in Eq.(\ref{xs-cs}) are,
\begin{eqnarray}
\nonumber 
F(gg\rightarrow {}^3D_1g)&=&\frac{16\alpha_s^3 \pi^2}
{81 M^3 s^2(M^2 - s)^5 (M^2 - t)^5 (s + t)^5}
\{102 M^{20} s^3 + 302 M^{20} s^2 t + 302 M^{20} s t^2 \\
\nonumber 
&+& 102 M^{20} t^3 - 286 M^{18} s^4 - 1732 M^{18} s^3 t - 2844 M^
18 s^2 t^2 - 1732 M^{18} s t^3 \\
\nonumber 
&-& 286 M^{18} t^4 +  275 M^{16} s^5 + 3840 M^{16} s^4 t 
   + 10289 M^{16} s^3 t^2 + 10289 M^{16} s^2 t^3 \\
\nonumber 
&+&3840 M^{16} s t^4 + 275 M^{16} t^5 - 227M^{14} s^6 
   - 5004 M^{14} s^5 t - 19569 M^{14} s^4 t^2 \\
\nonumber 
&-&29536 M^{14} s^3 t^3 - 19569 M^{14} s^2 t^4 
  - 5004 M^{14} s t^5 - 227 M^{14} t^6 + 410 M^{12} s^7 \\
\nonumber 
&+& 5137 M^{12} s^6 t + 23585 M^{12} s^5 t^2 
  + 47908 M^{12} s^4 t^3 + 47908 M^{12} s^3 t^4\\
\nonumber 
&+&23585 M^{12} s^2 t^5 + 5137 M^{12} s t^6 + 410 M^{12} t^7 
    - 470 M^{10} s^8 - 4220 M^{10} s^7 t \\
\nonumber 
&-&  19534 M^{10} s^6 t^2 - 47528 M^{10} s^5 t^3 - 63536 M^{10} s^4 t^4 
   - 47528 M^{10} s^3 t^5\\ 
\nonumber 
&-& 19534 M^{10} s^2 t^6 - 4220 M^{10} s t^7 
    - 470 M^{10} t^8 + 245 M^8 s^9 + 2190M^8 s^8 t\\ 
\nonumber 
&+& 10358 M^8 s^7 t^2 + 28602 M^8 s^6 t^3 
  + 47093 M^8 s^5 t^4 + 47093 M^8 s^4 t^5 \\
\nonumber 
&+& 28602 M^8 s^3 t^6 + 10358 M^8 s^2 t^7 + 2190 M^8 s t^8 
   + 245 M^8 t^9 - 49 M^6 s^{10} \\
\nonumber 
&-&  580 M^6 s^9 t - 2822 M^6 s^8t^2- 8984 M^6 s^7 t^3 
    -17653 M^6 s^6 t^4 - 21968 M^6 s^5 t^5 \\
\nonumber 
&-&  17653 M^6 s^4 t^6 - 8984 M^6 s^3 t^7
   -2822 M^6 s^2 t^8 - 580 M^6 s t^9 - 49 M^6 t^{10}\\
\nonumber 
&+&  67 M^4 s^{10} t + 210 M^4 s^9 t^2 
   +774 M^4 s^8 t^3 +2006 M^4 s^7 t^4 + 3147 M^4 s^6 t^5 \\
\nonumber 
&+& 3147 M^4 s^5 t^6 + 2006 M^4 s^4 t^7 
   + 774 M^4 s^3 t^8 + 210 M^4 s^2 t^9 + 67 M^4 s t^{10} \\
\nonumber 
&+& 25 M^2 s^{10} t^2 + 100 M^2 s^9 t^3 
   +220 M^2 s^8 t^4 + 340 M^2 s^7 t^5 + 390M^2 s^6 t^6 \\
\nonumber 
&+& 340 M^2 s^5 t^7 + 220 M^2 s^4 t^8 +
 100 M^2 s^3 t^9 + 25 M^2 s^2 t^{10} + 5 s^{10} t^3 \\
\nonumber
&+& 25 s^9 t^4 + 60 s^8 t^5 + 90 s^7 t^6 + 90 s^6 t^7 +
 60 s^5 t^8 + 25 s^4 t^9 + 5 s^3 t^{10}\},
\end{eqnarray}

\begin{eqnarray}
\nonumber 
F(gg\rightarrow {}^3D_2g)&=&\frac{32\alpha_s^3 \pi^2}
{27 M^3 s^2(M^2 - s)^5 (M^2 - t)^5 (s + t)^5}\{
  M^{20} s^3 + 9 M^{20} s^2 t + 9 M^{20} s t^2+M^{20} t^3 \\ 
\nonumber 
&-& 3 M^{18} s^4 - 64 M^{18} s^3 t 
   -130 M^{18} s^2 t^2 - 64M^{18} s t^3 -3 M^{18} t^4 + 5 M^{16} s^5 \\
\nonumber 
& +&  180 M^{16} s^4 t + 611 M^{16} s^3 t^2 
    +611 M^{16} s^2 t^3 + 180 M^{16} 
     s t^4 - 11 M^{14} s^6   \\
\nonumber  
&-& 280 M^{14} s^5 t - 1369 M^{14} s^4 t^2 - 2208 M^{14} s^3 t^3 
    -1369 M^{14} s^2 t^4 - 280 M^{14} s t^5\\ 
\nonumber 
&-&  11 M^{14} t^6 + 20 M^{12} s^7+ 269 M^{12} s^6 t + 1716 M^{12} s^5 t^2 
    +3927 M^{12} s^4 t^3 \\
\nonumber 
&+& 3927 M^{12} s^3 t^4 +1716 M^{12} s^2 t^5 + 269 
  M^{12} s t^6 + 20 M^{12} t^7 - 20 M^{10} s^8 \\
\nonumber 
&-& 1283 M^{10} s^6 t^2 - 3888 M^{10} s^5 t^3 - 5450 
  M^{10} s^4 t^4 - 3888 M^{10} s^3 t^5 \\
\nonumber 
&-&  1283 M^{10} s^2 t^6-144 M^{10} s t^7 - 20 M^{10} t^8 + 10 M^8 s^9 + 16 
     M^8 s^8 t+ 5 M^{16} t^5  \\
\nonumber 
&+& 568 M^8 s^7 t^2 + 2365 M^8 s^6 t^3 
   +4181 M^8 s^5 t^4 + 4181 M^8 s^4 t^5 + 2365 M^8 s^3 t^6  \\
\nonumber 
&+& 568 M^8 s^2 t^7 + 16 M^8 s t^8 + 10 M^8 t^9
 2 M^6 s^{10} + 20 M^6 s^9 t -144 M^{10} s^7 t \\
\nonumber 
&-& 156 M^6 s^8 t^2 - 1072 M^6 s^7 t^3 - 2230 M^6 s^6 t^4 
   - 2664 M^6 s^5 t^5 -2230 M^6 s^4 t^6 \\
\nonumber 
&-& 1072 M^6 s^3 t^7 - 156 M^6 s^2 t^8 
  + 20 M^6 s t^9 - 2 M^6 t^{10} - 6 M^4 s^{10} t \\
\nonumber 
&+& 50 M^4 s^9 t^2 + 472 M^4 s^8 t^3 + 1172 M^4 s^7 t^4 
  + 1570 M^4 s^6 t^5 + 1570 M^4 s^5 t^6\\ 
\nonumber 
&+& 1172 M^4 s^4 t^7 +472 M^4 s^3 t^8 + 50 M^4 s^2 t^9 - 6 M^4 s 
   t^{10} - 16 M^2 s^{10} t^2 \\
\nonumber 
&-& 160 M^2 s^9 t^3 - 496 M^2  s^8 t^4 -832 M^2 s^7 t^5 
   - 960 M^2 s^6 t^6 - 832 M^2 s^5 t^7 \\
\nonumber 
&-& 496 M^2 s^4 t^8 - 160 M^2 s^3 t^9 - 
  16 M^2 s^2 t^{10}+16 s^{10} t^3 + 80 s^9 t^4  \\
\nonumber 
&+& 192 s^8 t^5 + 288 s^7 t^6 + 288 s^6 t^7 + 192 s^5 t^8 
  80 s^4 t^9 + 16 s^3 t^{10}\},
\end{eqnarray}

\begin{eqnarray}
\nonumber
F(gg\rightarrow ^3D_3g)&=&\frac{256\alpha_s^3 \pi^2}
{189 M^3 s^2(M^2 - s)^5 (M^2 - t)^5 (s + t)^5}
\{8 M^{20} s^3 + 18 M^{20} s^2 t + 18 M^{20} s t^2 \\
\nonumber 
&+& 8 M^{20} t^3 - 24 M^{18} s^4 - 128 M^{18} s^3 t 
   -206 M^{18} s^2 t^2 - 128 M^{18} s t^3  \\
\nonumber  
&-& 24 M^{18} t^4 + 25 M^{16} s^5 + 300 M^{16} s^4 t 
  + 826 M^{16} s^3 t^2 +826 M^{16} s^2 t^3  \\
\nonumber 
&+&300 M^{16} s t^4 + 25 M^{16} t^5 - 13 M^{14} s^6 - 
   356 M^{14} s^5 t - 1556 M^{14} s^4 t^2 \\
\nonumber 
&-& 2424 M^{14} s^3 t^3 - 1556 M^{14} s^2 t^4 - 356 M^{14} s t^5 - 13 M^{14} 
t^6 + 10 M^{12} s^7 \\
\nonumber 
&+& 1680 M^{12} s^5 t^2 + 3717 M^{12} s^4 t^3 
  + 3717 M^{12} s^3 t^4 + 1680 M^{12} s^2 t^5 \\
\nonumber 
&+&  283 M^{12} s t^6 + 10 M^{12} t^7 -
10 M^{10} s^8 - 180 M^{10} s^7 t - 1201 M^{10} s^6 t^2 \\
\nonumber  
&-& 3342 M^{10} s^5 t^3 - 4624 M^{10} s^4 t^4 - 3342 M^{10} s^3 t^5 
    -1201 M^{10} s^2 t^6\\
\nonumber  
& -& 180 M^{10} s t^7 - 10 M^{10} 
  t^8 + 5 M^8 s^9 + 80 M^8 s^8 t + 602 M^8 s^7 t^2 \\
\nonumber 
&+&
 1943 M^8 s^6 t^3 + 3307 M^8 s^5 t^4 + 3307 M^8 s^
4 t^5 + 1943 M^8 s^3 t^6 \\
\nonumber 
&+& 602 M^8 s^2 t^7 + 80 M^8 s t^8 
   +5 M^8 t^9 - M^6 s^{10} - 20 M^6 s^9 t - 198 M^6 s^8 t^2 \\
\nonumber 
&-&776 M^6 s^7 t^3 - 1502 M^6 s^6 t^4 
    -1812 M^6 s^5 t^5 - 1502 M^6 s^4 t^6 - 776 M^6 s^3 t^7 \\
\nonumber 
& -&  198 M^6 s^2 t^8 - 20 M^6 s t^9 - M^6 t^{10} 
   +3 M^4 s^{10} t + 40 M^4 s^9 t^2 + 221 M^4 s^8 t^3 \\
\nonumber 
&+&  514 M^4 s^7 t^4 + 698 M^4 s^6 t^5 + 698 M^4 s^5 t^6 
    +514 M^4 s^4 t^7 + 221 M^4 s^3 t^8 \\
\nonumber 
&+&  3 M^4 s t^{10} - 5 M^2 s^{10} t^2 - 50 M^2 s^9 t^3 
    -155 M^2 s^8 t^4 - 260 M^2 s^7 t^5  \\
\nonumber 
&-& 300 M^2 s^6 t^6 - 260 M^2 s^5 t^7 - 155 M^2 s^4 t^8 - 50 M^2 s^3 t^9 \\
\nonumber 
&-& 
5 M^2 s^2 t^{10} + 5 s^{10} t^3
+ 25 s^9 t^4 + 60 s^8 t^5 + 90 s^7 t^6 + 90 s^6 t^7 \\
\nonumber 
&+& 60 s^5 t^8 + 25 s^4 t^9 + 5 s^3 t^{10}+ 283 M^{12} s^6 t \},
\end{eqnarray}
where $M=2m_Q$ and $s,~t,~u$ are Mandelsterm invariants for the processes
$gg\to {}^3D_Jg$.

\newpage
\centerline{\bf \large Figure Captions}
\vskip 1cm
\noindent
Fig.1. $D$-wave charmonium production in proton-nucleon collisions
for $x_F>0$.
The dashed lines are the color-singlet contributions and the
solid lines are color-octet contributions.
For the solid lines, from up to down, they are for $J=3,~2,~1$
$D$-wave sates respectively.
For the dashed lines, from up to down, they are for $J=3,~1,~2$
sates respectively.

\vskip 0.2cm
\noindent
Fig.2 $D$-wave bottomonium production in proton-nucleon collisions
for $x_F>0$. The lines are the same as those in Fig.1.
\vskip 0.2cm
\noindent
Fig.3 $D$-wave charmonium production in pion-nucleon collisions
for $x_F>0$. The lines are the same as those in Fig.1.
\vskip 0.2cm
\noindent
Fig.4 $D$-wave bottomonium production in pion-nucleon collisions
for $x_F>0$. The lines are the same as those in Fig.1.

\begin{figure}
\hbox{
\psfig{file=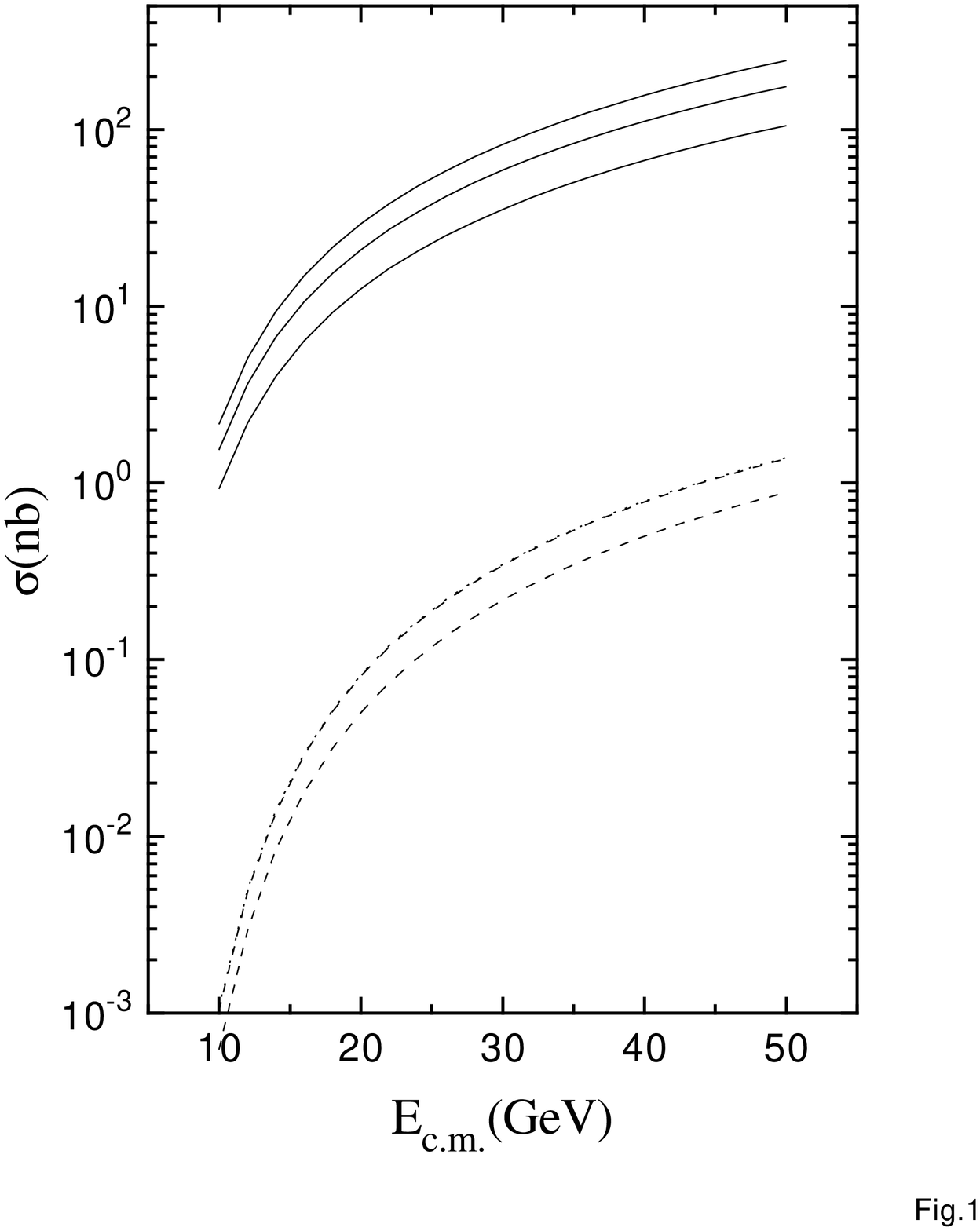,width=16cm,angle=0}}
\hbox{
\psfig{file=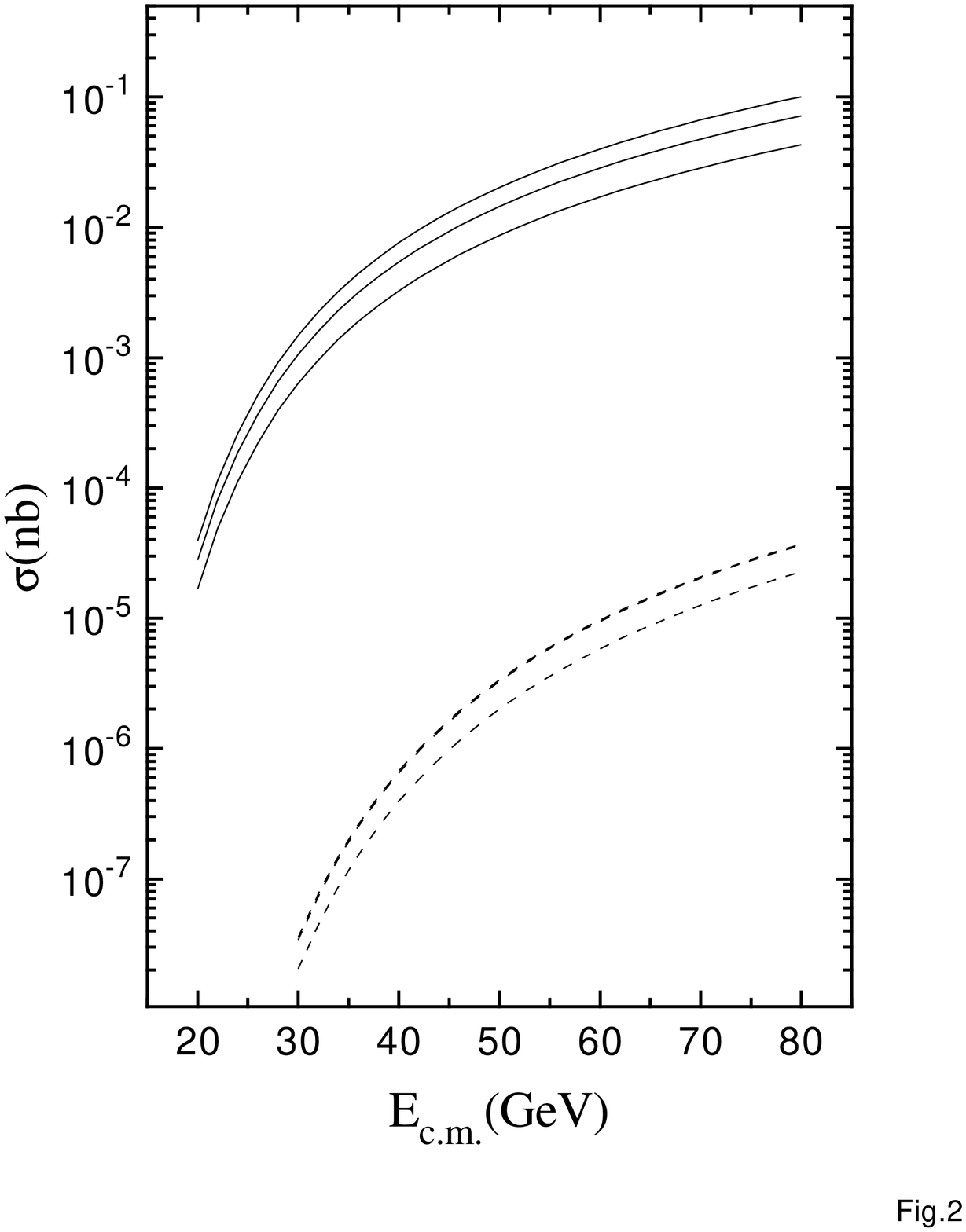,width=16cm,angle=0}}
\hbox{
\psfig{file=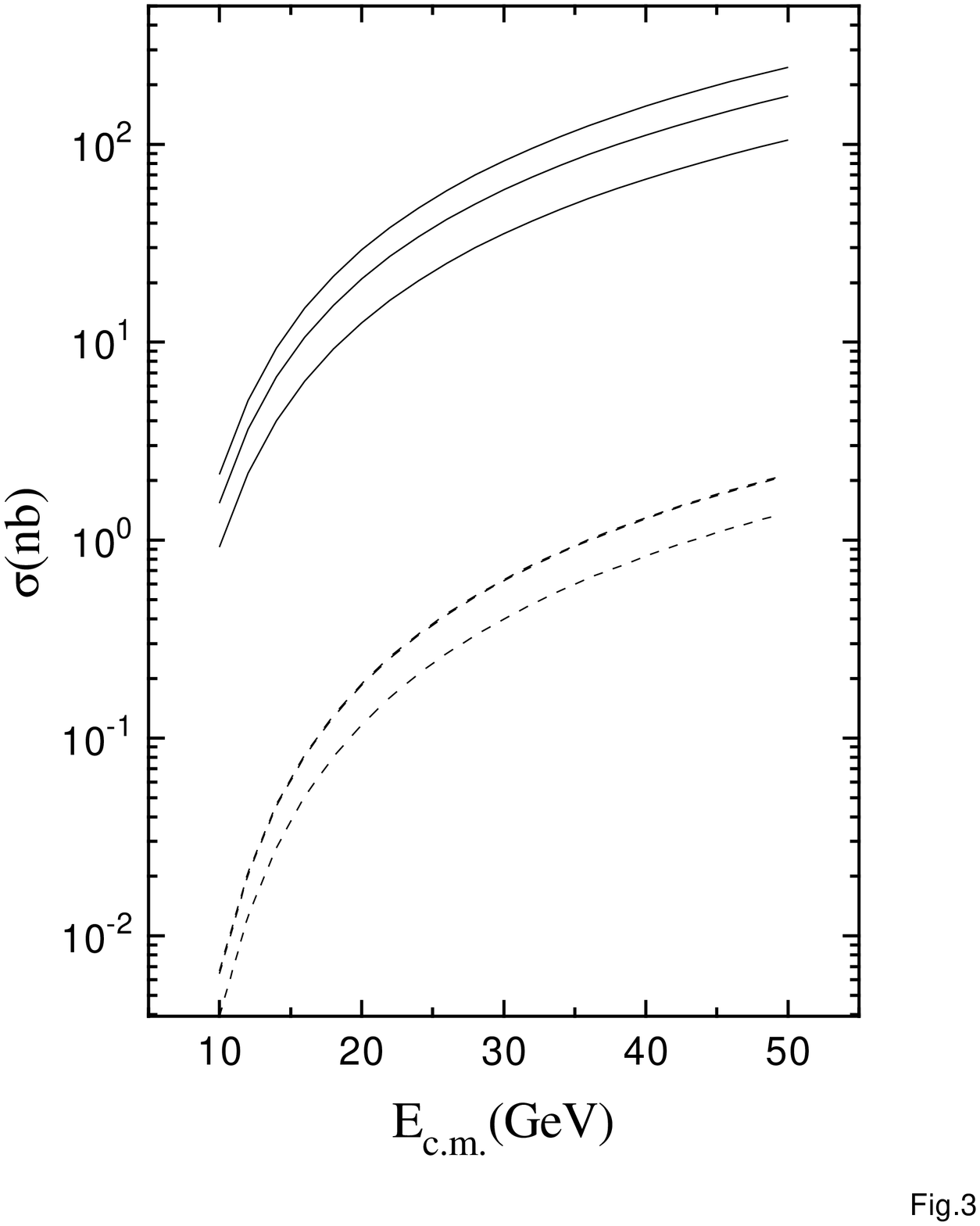,width=16cm,angle=0}
}
\hbox{
\psfig{file=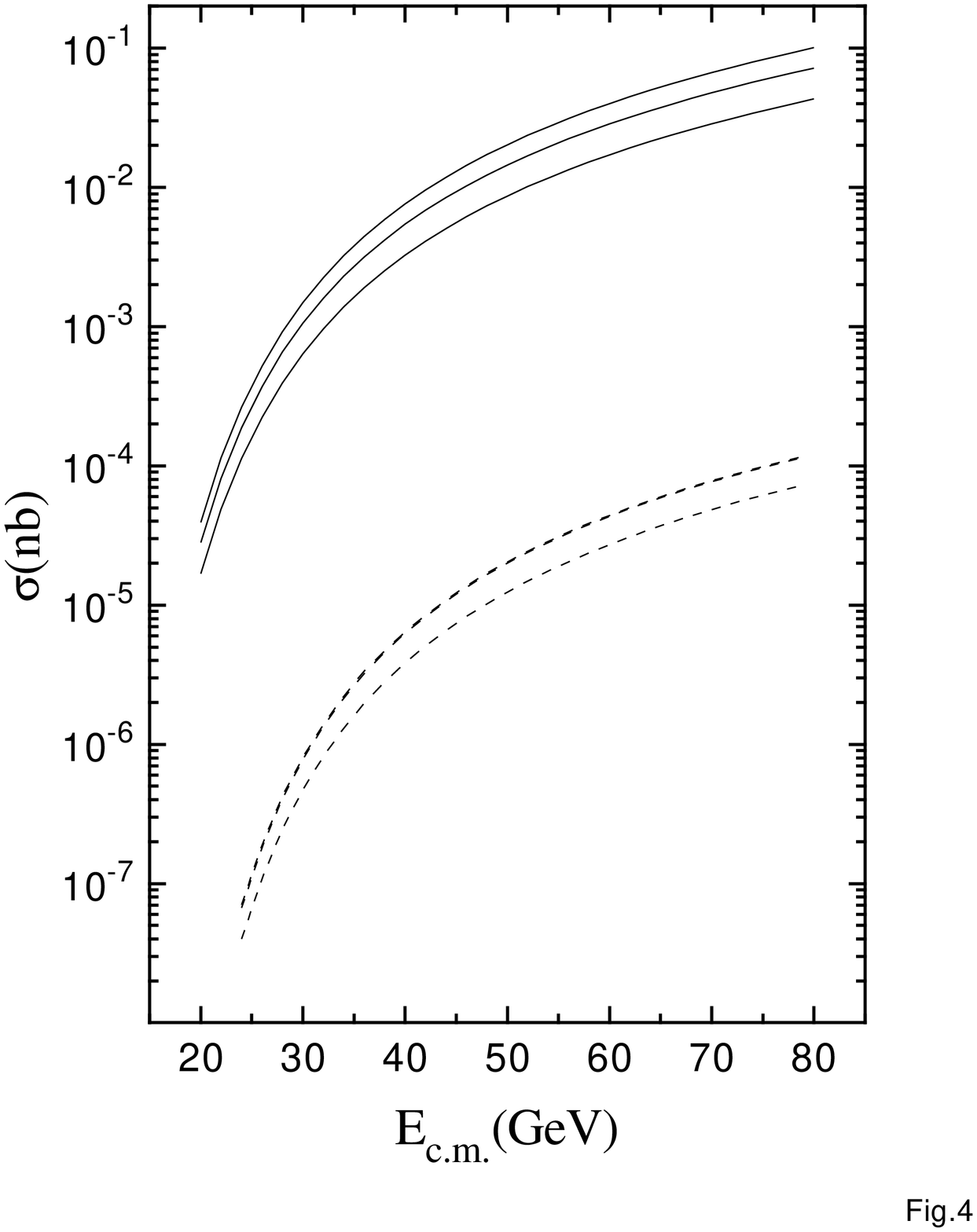,width=16cm,angle=0}
}
\end{figure}

\end{document}